\documentstyle[epsfig,referee]{mn}

\title{Stability of rotating spherical stellar systems} 
\author[J.-M. Alimi, J. Perez, A. Serna]{Jean-Michel Alimi$^{1}$, 
J\'er\^{o}me Perez $^{2}$, Arturo Serna$^{3}$ \\
$^{1}$ Laboratoire d'Astrophysique Extragalactique et de Cosmologie -CNRS 
URA 173 \\
Observatoire de Meudon - 5, Place Jules Jansen - 92195 Meudon 
-France\\
$^{2}$ Ecole Nationale Sup\'erieure des Techniques Avanc\'ees - 
 SMP - CNRS URA D853\\
32 Boulevard Victor - 75996 Paris - 
France\\
$^{3}$ Universidad Miguel Hern\'andez, Area de F\'{\i}sica Aplicada,
Edif. La Galia, 03291-Elche, Spain} 
\date{Accepted 199- Received 1995 in original form 199-} 
\pubyear{1997}
\begin{document}

\maketitle
\begin{abstract}
  We study  the stability  of rotating collisionless  self-gravitating
  spherical  systems by using high  resolution N-body experiments on a
  Connection Machine CM-5.
  
  We added  rotation  to  Ossipkov-Merritt (hereafter OM)  anisotropic
  spherical systems by using two methods.   The first method conserves
  the anisotropy  of  the  distribution function   defined in the   OM
  algorithm.  The second  method distorts   the systems in    velocity
  space.  We  then show that the stability  of systems depends both on
  their anisotropy and on  the  value of the  ratio between  the total
  kinetic energy and the rotational kinetic energy.   We also test the
  relevance of the  stability  parameters introduced  by Perez et  al.
  (1996) for the case of rotating systems.

\end{abstract}
\section{Introduction}
Various         analytical       and         numerical         studies
\cite{antonov73,barnes86,palmer86,perez1,perez2}     (and   references
therein)  have  shown that  spherical, collisionless, self-gravitating
anisotropic systems with components moving mainly on radial orbits are
unstable. However,  all these  works considered  non-rotating systems,
and it is well known that  rotation can play  an important role in the
dynamical    evolution  of  systems    and   modify  their   stability
properties. It has been  shown that rotation  can be the cause of
the deformation of systems like globular clusters or weakly elliptical
galaxies \cite{goodwin97,stavena96}.

The stability of  rotating stellar systems  is a very complex problem,
Much work has been  concerned with barred  galaxies (which are rapidly
rotating stellar systems) (for a review see, IAU Colloquium 157 ), but
in a  more   general context, few  studies  have  been devoted in  the
literature to this topic.  For  example, Papaloizou, Palmer and  Allen
(1991) have performed a series of numerical simulations to analyze the
stability   of  systems where  rotation  was  introduced  by using the
technique proposed by   Lynden-Bell  (1962).  All   their  simulations
produced endstates in  which  a triaxial bar appears.  These important
results cannot  be considered  general, since  they  were obtained for
systems dominated by particles evolving on  radial orbits, and was put
in  rotation  by  a  specific  procedure.   In   order to  analyze the
influence of rotation on  the (in)stability of  a given system, it  is
necessary to consider not only  spherical systems with different kinds
of anisotropy but also different methods for introducing the rotation.
This   paper develops such  an  analysis.   We are  also interested in
testing the relevance  of  the stability  parameters  \cite{perez2} on
rotating systems.  Perez et al.  (1996) have  shown that the stability
of spherical self-gravitating non-rotating systems can be deduced from
the  'anisotropic'   component of   the     linear variation   of  the
distribution function   (see  below   Section  2.2).   Such  stability
parameters  can be computed from  rotating systems.  We show that they
are still relevant  for anisotropic systems as  long as the rotational
kinetic energy is not too large.

The paper is arranged as follows. We describe in  Section 2 the method
that we use to obtain the initial non-rotating  systems as well as the
parameters describing the (in)stability of such systems. In Section 3,
we detail the techniques used  to introduce a parameterizable rotation
to the initial  conditions presented in Section 2.   In Section  4, we
show our numerical results  on  the (in)stability of various  rotating
systems generated with different  procedures.  Finally, the discussion
and physical interpretation of our results are presented in Section 5.

\section{Stability and Instability of Non-rotating Systems}
\subsection{Non-rotating Initial Conditions}

In  a   previous  paper   \cite{perez2},  we used  the    OM algorithm
\cite{osipkov,merritt1,merritt2,bt}       to     generate  anisotropic
self-gravitating spherical  systems with  various physical properties.
This  algorithm    starts  from  a      density given by    $\rho_{\rm
iso}(r)\propto \psi_{\rm iso}^n$, where $\psi_{\rm iso}(r)$ is a known
gravitational potential satisfying  the Poisson equation, while $n$ is
the  polytropic index  ($1/2 <   n \leq   5$).  This  density  profile
$\rho_{iso}(r)$ is then deformed according to:

\begin{equation}
\rho_{\rm ani}(r) \;\;:=\;\;\left 
( 1+\frac{r^{2}}{r_{\rm a}^{2}} \right ) 
\rho_{\rm iso}(r) \, ,
\label{rhoani}
\end{equation}

\noindent
where the anisotropic radius $r_{\rm a}$ is a parameter which controls
the deformation.

Using the Abel  inversion technique,  this   procedure allows one   to
define an  anisotropic   distribution  function  (hereafter  DF) which
depends    both   on  $E$     and  $L^{2}$   through      the variable
$Q\;\;:=\;\;E+L^{2}/2r_{\rm a}^{2}$

\begin{eqnarray}
 f_{o}(Q)\;\;=\;\;\frac{\sqrt{2}}{4\pi^{2}}\frac{d}{dQ}\int_{Q}^{0}
 \frac{d\psi_{\rm   iso}}{\sqrt{\psi_{\rm   iso}-Q}}\;\frac{d\rho_{\rm
 ani}} {d\psi_{\rm iso}} \, .
\label{df_ossipkov_merritt}
\end{eqnarray}

Once this DF  has been computed,  the initial conditions of our N-body
numerical simulations are generated by  choosing  at random, from  the
above DF,  the positions  and velocities  for  the $N$ particles.  The
density profile  $\rho_{ani}(r)$  defined by equation  \ref{rhoani} is
the probability density from which   the positions are generated.  The
velocities are generated from the velocity probability density deduced
from the equation \ref{df_ossipkov_merritt} (see appendix).

It must be  noted that, there is   a fundamental limitation in the  OM
models: Any given value of the polytropic index $n$ implies a critical
value of $r_a$  below which the DF  becomes negative and unphysical in
some  region of   phase   space.   Merritt  (1985a)  interprets   this
limitation as a simple illustration of the well-known fact that radial
orbits   cannot  always   reproduce    an arbitrary   spherical   mass
distribution.  In theses cases, in order to extend the OM algorithm to
highly radially   anisotropic   ($r_a  \simeq 1$)   systems,  we  have
arbitrarily set the DF equal to zero in this region.  Such a procedure
on DF affect only particles with a large value of $Q$.  This procedure
is  applied for systems  with a  small value of  $r_a$ which  contains
mainly particles  with a small value  of $Q$.  Such a procedure affect
then a very small number of particles (less  than $0.1\%$ of the total
number of particles).  The systems with a modified DF are not strictly
OM systems.  However  they conserve the  properties which are for  the
present work:  the density profils deduced   from the modified  DF are
indistinguishable from the density profils  given by equation (1) with
the same value of $r_a$, the Lindblad diagrams  are very peaked around
a small value  of $Q$ \cite{perez2},   they well correspond  to highly
radially anisotropic systems, and finally the radial dependence of the
velocity anisotropy
\begin{equation} 
  \frac{\sigma_{r}^{2}}{\sigma_{t}^{2}}\;\;
:=\;\;\frac{<v_{r}^{2}>}{\frac{1}                                  {2}
<v_{t}^{2}>}\;\;=\;\;1+\frac{r^{2}}{r_{a}^{2}} \, . 
\end{equation}
is preserved.

Finally,  since  each   particle   is initialized independently,   the
equilibrium DF  $f_{o}(E,L^{2})$ of the   system  is in fact  slightly
perturbed. The perturbation is due to local Poissonian fluctuations of
the density. The dynamical evolution of the system then represents the
response of  an   anisotropic  self-gravitating spherical  equilibrium
system submitted to such a perturbation.

\subsection{Stability Analysis}

The equilibrium DF of a spherical self-gravitating system depends only
on the one-particle energy $E$ and the  squared total angular momentum
$L^{2 }$. If  $g_{1}$ denotes the  perturbation generator,  the linear
variation of the DF can be written as

\begin{eqnarray}
\delta f\;=\;\frac{\partial f_{o}}{\partial E}\{g_{1},E\}+ 
\frac{\partial f_{o}}{\partial L^{2}}\{g_{1},L^{2}\}.
\label{deltaf}
\end{eqnarray}

If DF   is  a monotonic decreasing  function\footnote{We  consider all
along this paper only   systems with a  DF  which admits  a monotonous
decreasing  dependence with respect to all  the isolating integrals of
motion ($\frac{\partial  f_{o}}{\partial E} <  0$, and $\frac{\partial
f_{o}}{\partial  L^{2}} < 0$).}, the  stability is then related to the
Poisson  brackets     $\{g_{1},E\}$       and        $\{g_{1},L^{2}\}$
\cite{perez1,perez2}.    In our N-body  simulations,  these quantities
appear as two random variables, $\epsilon$ and $\lambda$ respectively,
defined  for  each particle $i$  \cite{perez2}.   The stability of the
system can be predicted    from  the probability $P_{\epsilon}$    for
$\epsilon$  to    be negative, and    the   statistical Pearson  index
\cite{stat} $P_{\lambda}$ of the  variable $\lambda$.  All anisotropic
collisionless self-gravitating non-rotating spherical systems with
\begin{equation}
P_{\lambda} \la 2.5 \; \mbox{and} \; P_{\epsilon} \ga 20\%
\label{crit_unstable}
\end{equation}
are unstable, while those with
\begin{equation}
P_{\lambda} \ga 2.5 \; \mbox{and} \; P_{\epsilon} \la 20\%
\label{crit_stable}
\end{equation}
are stable. The two other  regions of the $(P_{\lambda} P_{\epsilon})$
plane, correspond  to a  transition between  a  stable system  and  an
unstable  system. In the particular  case of  OM models, these regions
correspond to  an    anisotropy  radius  $r_{\rm    a}$ close to     2
\cite{perez2}.

\section{Generation of Rotating Systems}
\subsection{Definition}
In order to   generate virial-relaxed rotating spherical   systems, we
modify  the non-rotating systems  defined  in the previous section  by
using techniques derived  from  the Lynden-Bell method  (1962).  Since
this method preserves  the  position and  the norm of  the velocity of
each particle, the systems are put in rotation without modifying their
total  potential and kinetic   energy.    In practice,  we   apply the
following   transformations   to   the  velocity  components   $\{v_r,
v_{\theta}, v_{\phi}\}$ of the particles:

\begin{equation}
\begin{array}{ll}
 Method \; 1 & Method \; 2 \\
v_{r}\longrightarrow v_{r} & v_{r}\longrightarrow 0. \\
v_{\theta}\longrightarrow v_{\theta} & v_{\theta}\longrightarrow 
v_{\theta} \\
v_{\phi}\longrightarrow\mid\!v_{\phi}\!\mid & v_{\phi}\longrightarrow 
\displaystyle{ \sqrt{1+\frac{v_r^2}{v_{\phi}^2}} } 
\mid\!v_{\phi}\!\mid
\end{array}
\end{equation}

The first method  then conserves the radial  anisotropy defined in the
OM algorithm, while the second  method distorts the system in velocity
space.

The amount of rotation introduced  by these  methods can be  evaluated
through the ratio:
\begin{eqnarray}
\mu = K_{rot} / K_{tot},
\label{mu}
\end{eqnarray}

\noindent
where $K_{tot}$ is the total  kinetic energy,  and $K_{rot}$ is  the
rotation kinetic energy defined by Navarro and White (1993):

\begin{equation}
K_{rot}=\frac{1}{2}\sum_{i=1}^{N} m_i \frac {(\mbox{\bf \sl L}_i \cdot 
\hat{\mbox{\bf \sl L}}_{tot})^2} {[r_{i}^{2}-(\mbox{\bf \sl r}_i \cdot 
\hat{\mbox{\bf \sl L}}_{tot})^2]}
\label{krot}
\end{equation}
Here, {\bf \sl L}$_i$ is the specific angular  momentum of particle i,
$\hat{\mbox{\bf  \sl L}}_{tot}$ is a unit   vector in the direction of
the  total angular momentum  of the system, when  the  system does not
rotate  at all the $\hat{\mbox{\bf \sl   L}}_{tot}$ vector is the null
vector, and $N$ is the total number of particles.  In order to exclude
counter-rotating  particles,   the sum   in  equation (\ref{krot})  is
actually    carried out   only  over those    particles satisfying the
condition $(\mbox{\bf \sl  L}_i \cdot \hat{\mbox{\bf \sl  L}}_{tot}) >
0$.

In order to have systems with different strengths of {\it homogeneous}
rotation  (HR), we   have applied either  Method 1   or Method 2  to a
fraction $\tau$  of the total  number of  particles. This fraction has
been constructed by  choosing the particles at  random  in the overall
system. When $\tau\rightarrow0$, the  system does not turn while, when
$\tau\rightarrow1$, the system rotation reaches its maximum value.

In  principle, there is   no reason   to  consider  only the  case  of
homogeneous rotation.   Moreover, in   order   to roughly  model   the
presence of a rotating massive object  like those sometimes considered
in  the center  of  some elliptical galaxies,  we  also  consider {\it
inhomogeneous} rotations   (IR).   In this  case,  we  apply the above
velocity transformations only  to those particles  placed at a  radial
distance smaller  than  $k  \times  R_{\frac{1}{2}}$, where  $k$  is a
positive parameter and $R_{\frac{1}{2}}$ is the radius containing half
of the  system mass.  If $k=0$,   the system does  not  turn while, if
$k\rightarrow  + \infty$ the rotation has  its  maximum value. We have
then  four possible procedures  to introduce a  rotation motion on our
initial  conditions.    The   first  two  possibilities  introduce   a
homogeneous   rotation by choosing particles  at   random in the whole
system and modifying their velocities according either to the method 1
or 2, what defines the HR$_1$ and HR$_2$ procedures, respectively. The
other   two possibilities   introduce an  inhomogeneous   rotation  by
applying either  the method 1 or 2  to modify  the velocities of those
particles placed  within a given  radial  distance, which  defines the
procedures IR$_1$ and IR$_2$, respectively.

\begin{figure}
\centerline{\epsfig{file=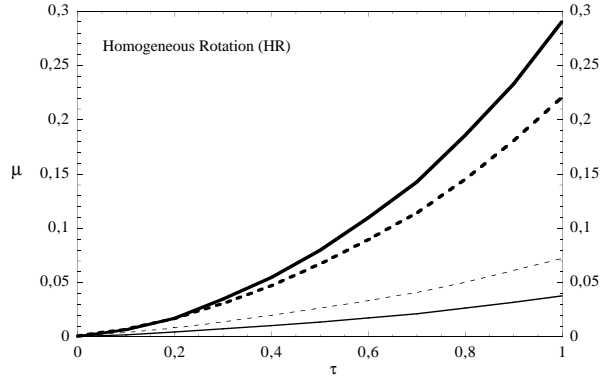,width=8cm} }
\caption{
 $\mu$ vs $\tau$ for $n=4$; HR$_{1}$,$r_{a}=1.5$ (thin solid line); 
 HR$_{1}$, $r_{a}=100$ (thin dashed line) and HR$_{2}$,$r_{a}=1.5$ (bold 
 Solid line); HR$_{2}$, $r_{\rm a}=100$ (bold dashed line)
 \label{krothom} }
\end{figure}

\begin{figure}
\centering
\epsfig{file=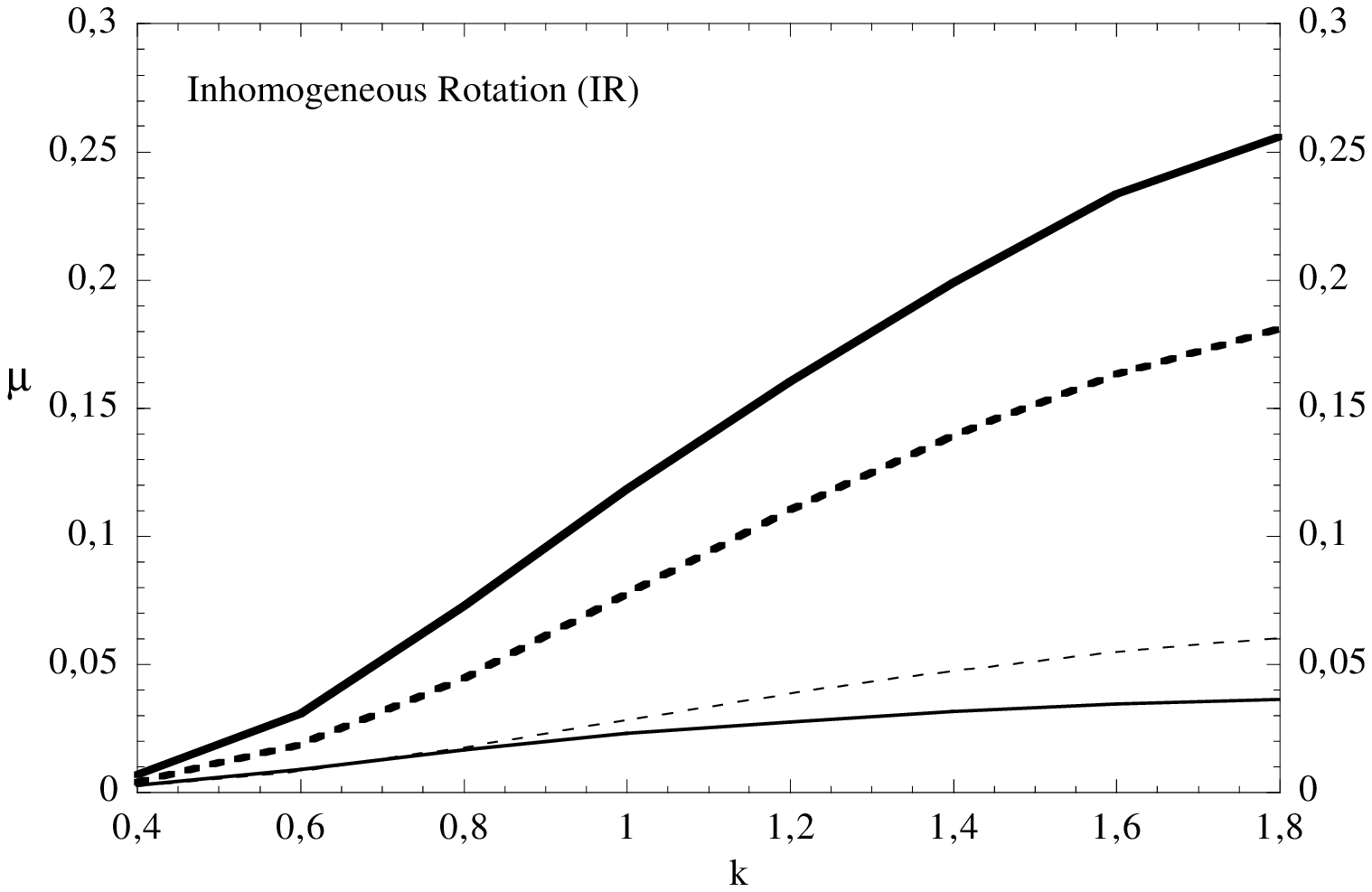,width=8cm} 
\caption{
\label{krotinh}
$\mu$ vs  $k$ for $n=4$,  IR$_{1}$, $r_{\rm a}=1.5$ (thin solid line);
IR$_{1}$, $r_{\rm a}=100$ (thin   dashed line) and   IR$_{2}$, $r_{\rm
  a}=1.5$ (bold  Solid  line); IR$_{2}$, $r_{\rm a}=100$  (bold dashed
line) }
\end{figure}

Figures \ref{krothom} and   \ref{krotinh}  show the values    of $\mu$
(equation \ref{mu}) obtained from the four  possible procedures. As we
can see from such figures, only the HR$_2$  and IR$_2$ procedures lead
to large fractions  ($\mu\geq 10\%$) of  kinetic rotation energy.   We
also note from these  figures that  the dependence  of $\mu$  on $r_a$
depends  on  whether velocities have  been   modified according to the
method 1 or 2.   As a matter  of fact,  for the HR$_{1}$  and IR$_{1}$
procedures,   the amount of  rotation  obtained ($\mu$) is greater for
large $r_a$ values than for small ones.  On the contrary, for HR$_{2}$
and IR$_{2}$, $\mu$  is larger for  small $r_a$ values  than for large
ones. 

\section{Influence of the Rotation on the Stability}

Using the N-body  code described in  Alimi and Scholl (1993), we  have
performed  on   Connection-Machine    5   a    series of     numerical
simulations\footnote{The set of  numerical  simulations performed have
been made with 16384  particles.  Some experiments have been performed
using more particles (65536), no significant change in comparison with
the work presented here have been  obtained.}  of the evolution of the
systems    defined in the    previous  section.  As the  collisionless
hypothesis is fundamental  for interpreting our  results,  we have not
continued  our simulations beyond  a  few  hundred dynamical times  in
order  to avoid the later  evolution where two-body relaxation arises.
However, all  our models reach  a steady  state before about  $50~T_d$
(where  the  initial  dynamical time  is  estimated  by the  following
formula  $T_d  =  \sqrt{\sum r^2_i/ \sum   v^2_i}$,  the summations on
initial positions and velocities  are done on all  the particles).  We
will then present our results for this interval.

The physical   mechanism of      the radial-orbit   instability    for
collisionless self-gravitating systems  is  well known.  It has   been
described  in detail by     several authors (see  \cite{palmer4}   for
example).  The morphological deformation  of the initial gravitational
system resulting  from this instability is  mainly due to the trapping
of particles  with  a  low  angular  momentum  in a  bounded  area  of
space.  This trapping favors   a deformation of the initial  spherical
system  to an ellipsoidal or  even  a bar-like structure.  To evaluate
such a  deformation, it is convenient to  use the  axial ratio defined
from the moment of inertia tensor $I$  \cite{allen90}.  From the three
real eigenvalues of $I$, $\lambda_1 \geq \lambda_2 \geq \lambda_3$, we
compute the axial   ratios $a_1 =  \lambda_2 /  \lambda_1$  and $a_2 =
\lambda_2 /  \lambda_3$.  These two  quantities,  which can always  be
defined  because these eigenvalues  never  vanish, satisfy $a_1 \leq 1
\leq a_2$.   In order  to  discriminate   clearly between  a  bar-like
structure, a  quasi-sphere and a disk-like  structure ,  we define the
quantity $f$ from $a_1$ and $a_2$
\begin{equation}
f = \frac{1 - a_1}{a_2 - 1}
\label{f}
\end{equation}
A bar-like structure is characterized by $a_1  < 1$ and $a_2 \simeq 1$
which  implies a $f$ value significantly  larger than $1$. A disc-like
structure is characterized by $a_1 \simeq 1$, $a_2 > 1$ and  $ 0 < f <
1$.  Any system with a $f$ value  of order unity has a quasi-spherical
structure.

\subsection{Rotating systems according to Method 1}

\begin{figure*}
\centerline{ \epsfig{file=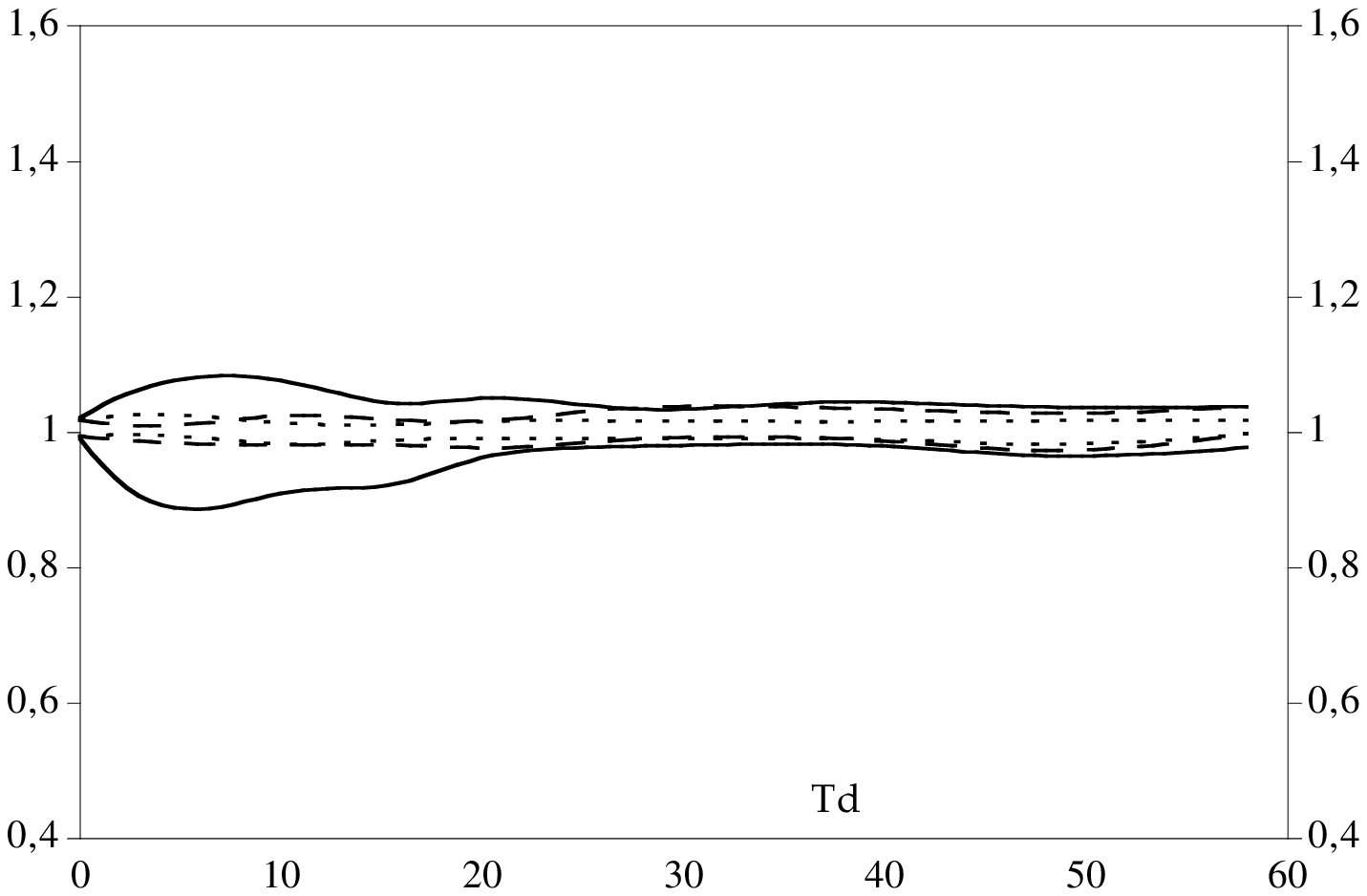,width=6cm}
\epsfig{file=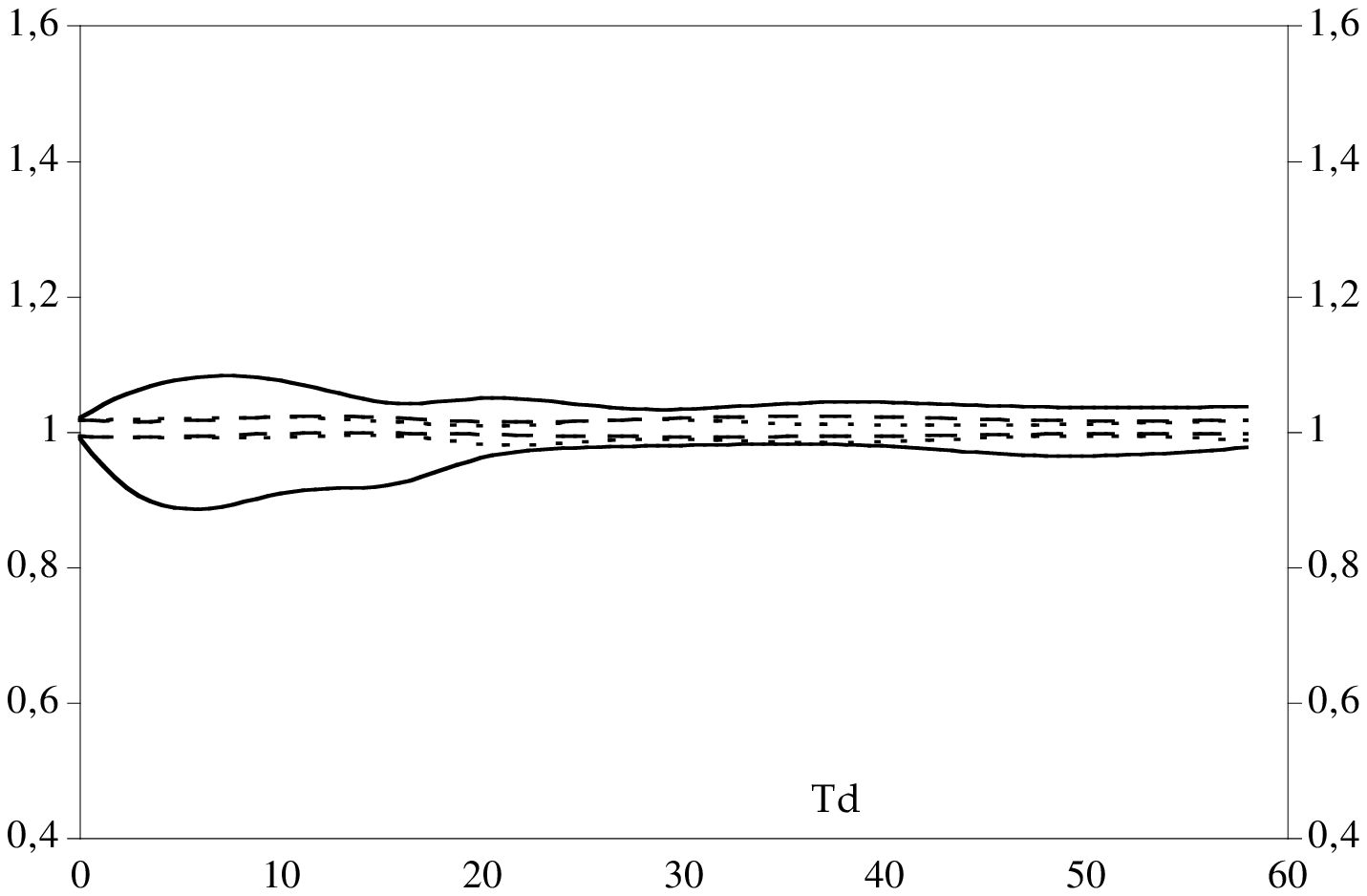,width=6cm} } \hfill \break \centerline{
\epsfig{file=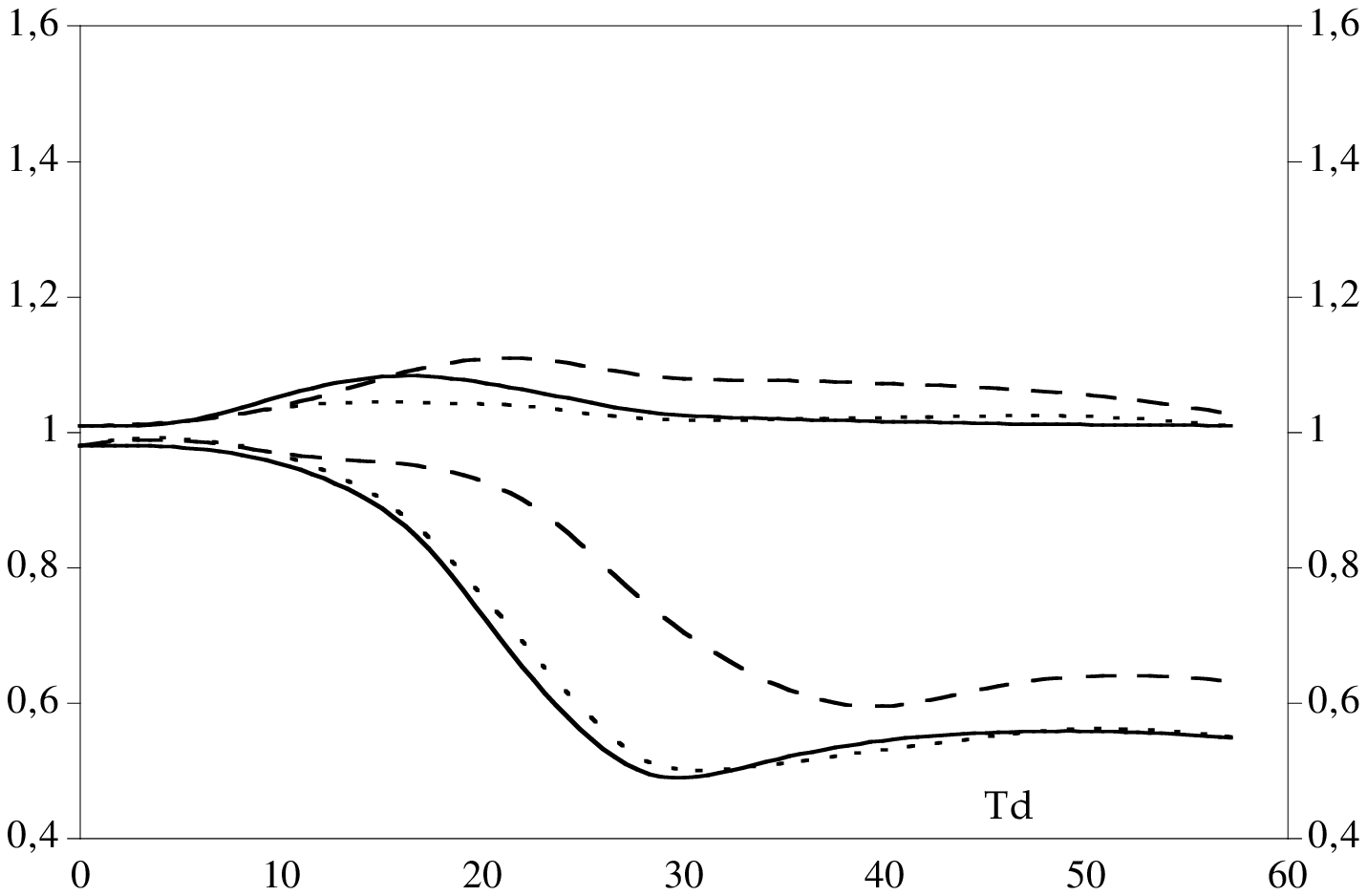,width=6cm} \epsfig{file=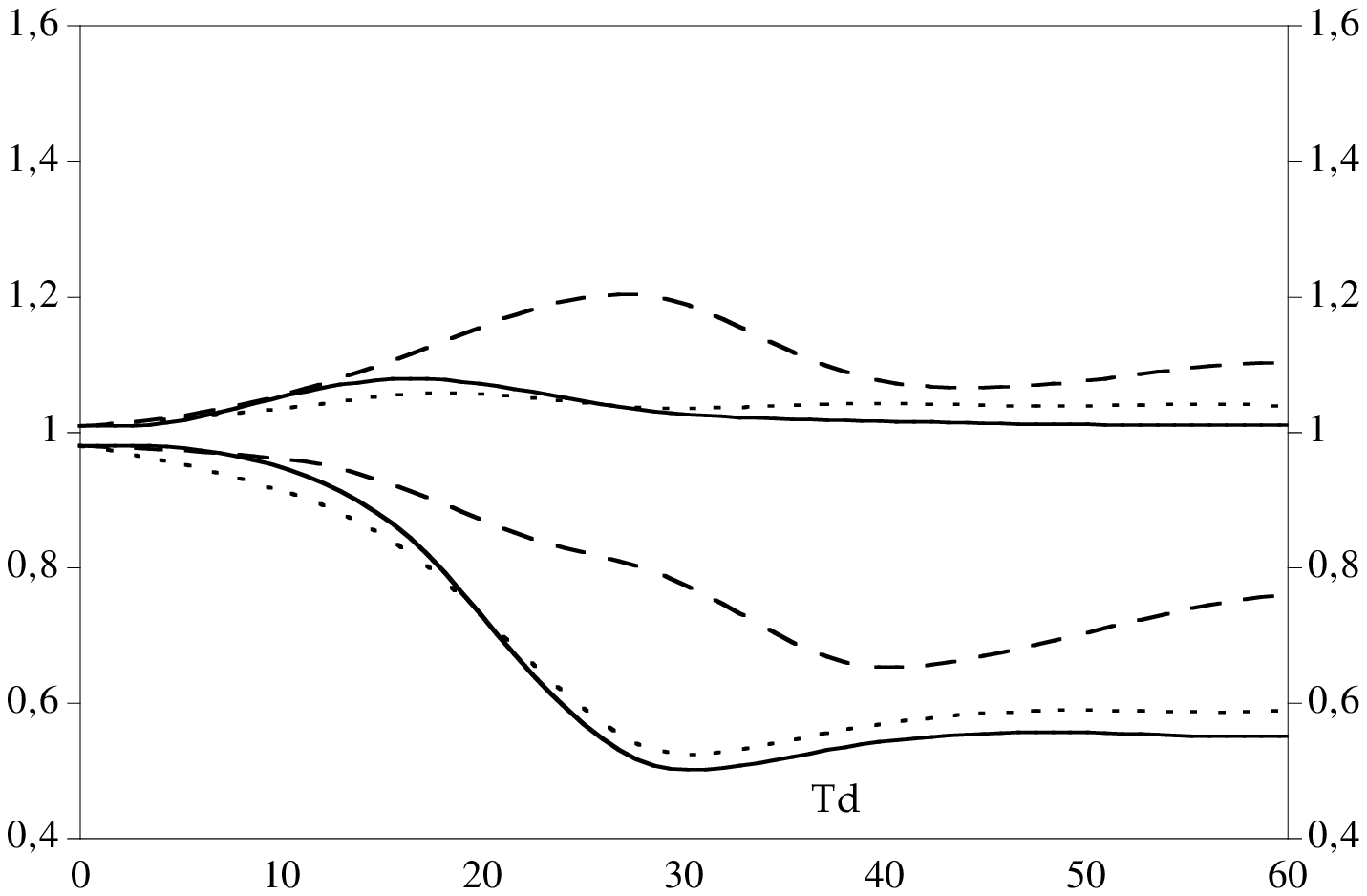,width=6cm}}
\caption{ The axial ratio vs. dynamical time for $n=4$; 
 $r_{\rm a}=100$-HR$_{1}$ (top-left panel), $r_{\rm a}=100$-IR$_{1}$
 (top-right panel), $r_{\rm a}=1.5$-HR$_{1}$ (bottom-left panel) and
 $r_{\rm a}=1.5$-IR$_{1}$ (bottom-right panel). The amount of
 rotation is represented by using different kinds of lines. Solid lines:
 $\tau=0$(HR) or $k=0$ (IR), dotted lines : $\tau=0.3$ (HR) or $k=0.6$ (IR)
 and dashed lines : $\tau=0.8$ (HR) or $k=1.4$ (IR). \label{ratio1} }
\end{figure*}

This type of rotation preserves the anisotropy  of the non-rotating OM
systems.  The distribution function   of the rotating  system  depends
only on the variable $Q\;=\;E+L^{2}/2r_{\rm a}^{2}$ as  in the case of
the non-rotating systems.  We see in Table  1 that, in the case $n=4$,
the  stability parameters defined   in   Section 2.2 are  very  weakly
modified whatever the $\tau$ and $k$ parameters values are, that is to
say, whatever the rotational kinetic energy is (low with this method).
According to the conditions given by equations \ref{crit_unstable} and
\ref{crit_stable},  we expect the     (in)stability of  the   original
non-rotating systems not to be modified when they are put in rotation.
Our numerical simulations confirm this. In figure \ref{ratio1}, we see
that the evolution of axial ratios is similar for the rotating (dashed
and dotted lines) and non-rotating (solid lines) systems. This results
holds for the homogeneous and inhomogeneous rotations and whatever the
$n$ parameter value is.
\begin{table*}
\label{stabparam1}
\begin{tabular}{||c|cc|cc||c|cc|cc||} \hline
HR$_{1}$ & & & & & IR$_{1}$ & & & & \\%
& $r_{\rm a}=1.5$ & & $r_{\rm a}=100$ & & &$r_{\rm a}=1.5$ & & 
$r_{\rm a}=100$ &\\ 
\hline
$\tau $ & $P_{\varepsilon}$ & $P_{\lambda}$ & 
 $P_{\varepsilon}$ & $P_{\lambda}$ &
$k $ & $P_{\varepsilon}$ & $P_{\lambda}$ & 
 $P_{\varepsilon}$ & $P_{\lambda}$ \\ 
\hline
0.0&22.42&1.39&13.84&4.07&0.4&24.54&1.31&15.25&3.89\\
0.2&22.83&1.41&14.28&4.12&0.6&24.04&1.31&14.60&3.84\\
0.4&22.76&1.39&14.05&4.23&0.8&24.02&1.33&14.44&3.75\\ 
0.6&22.22&1.41&13.68&4.22&1.0&23.57&1.27&17.18&3.77\\
0.8&22.04&1.42&13.34&4.08&1.2&22.04&1.25&14.09&3.79\\
1.0&21.93&1.44&12.98&4.18&1.4&23.59&1.26&13.99&3.79\\ \hline
\end{tabular}
\caption{Stability parameter evolution for the first method 
for rotating systems with $n=4$}
\end{table*}

\subsection{Rotating systems according to the second method}
\begin{figure}
\centerline{ \epsfig{file=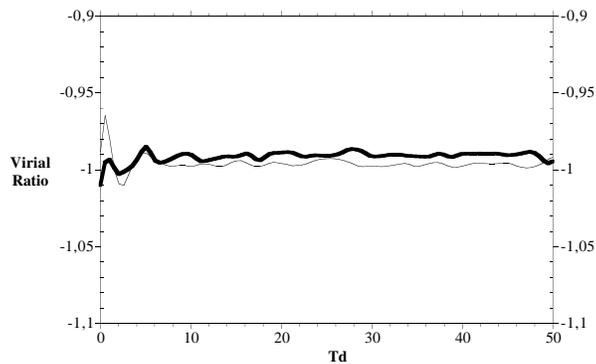,width=9cm}}
\caption{The evolution of  the Virial ratio   for the initial  systems
defined respectively by $n=4$,  $r_a=1.5$, and $n=4$, $r_a=100$  which
have  acquired their rotation according to   the procedure HR$_2$ with
$\tau=30\%$  (thin  line) and  IR$_2$   with $k=1.4$  (bold line) .  A
similar evolution for the Virial ratio (remaining very  close to 1) is
obtained for all runs.  \label{viriel} } \hfill \break
\end{figure}

The situation is now  more complicated because the  rotation procedure
modifies the system's anisotropy. In the second method, the stationary
OM  distribution  function is  modified   by  a positive definite  and
time-independant transformation.   The  resulting  DF  is then  always
stationary and positive  definite.   Moreover as the  modified systems
are  always   spherical (no  modification   on  positions   have  been
performed), the new DF depends only  on isolating integrals of motion,
the energy  and the  squared  angular momentum \cite{perez1}.   It  is
therefore a stationary solution of the collisionless Boltzmann-Poisson
system.  We have also  verified that the  resulting systems are always
virialized, as confirmed by figure \ref{viriel}.
 
\begin{figure*}
\centerline{\epsfig{file=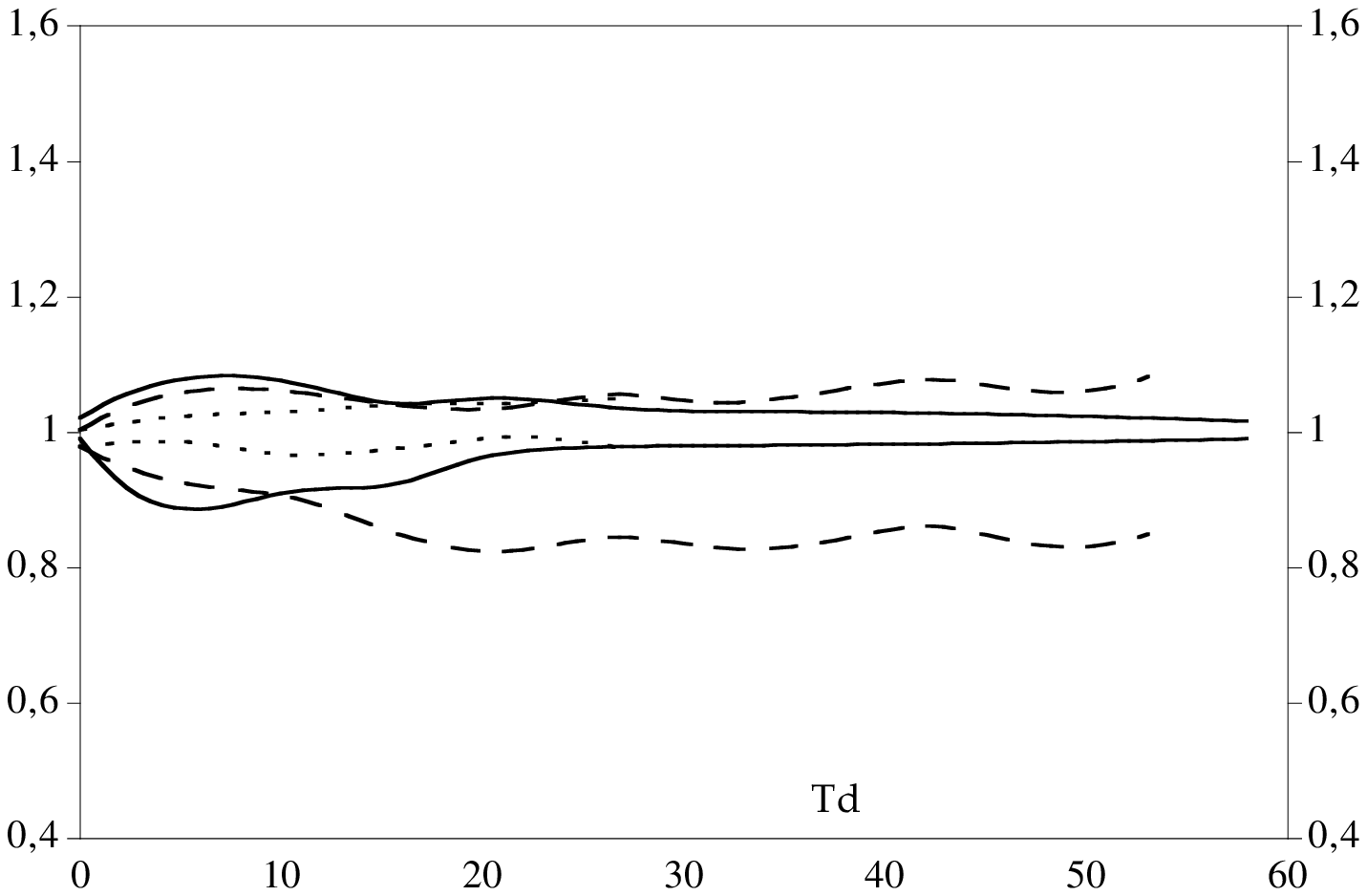,width=6cm}
\epsfig{file=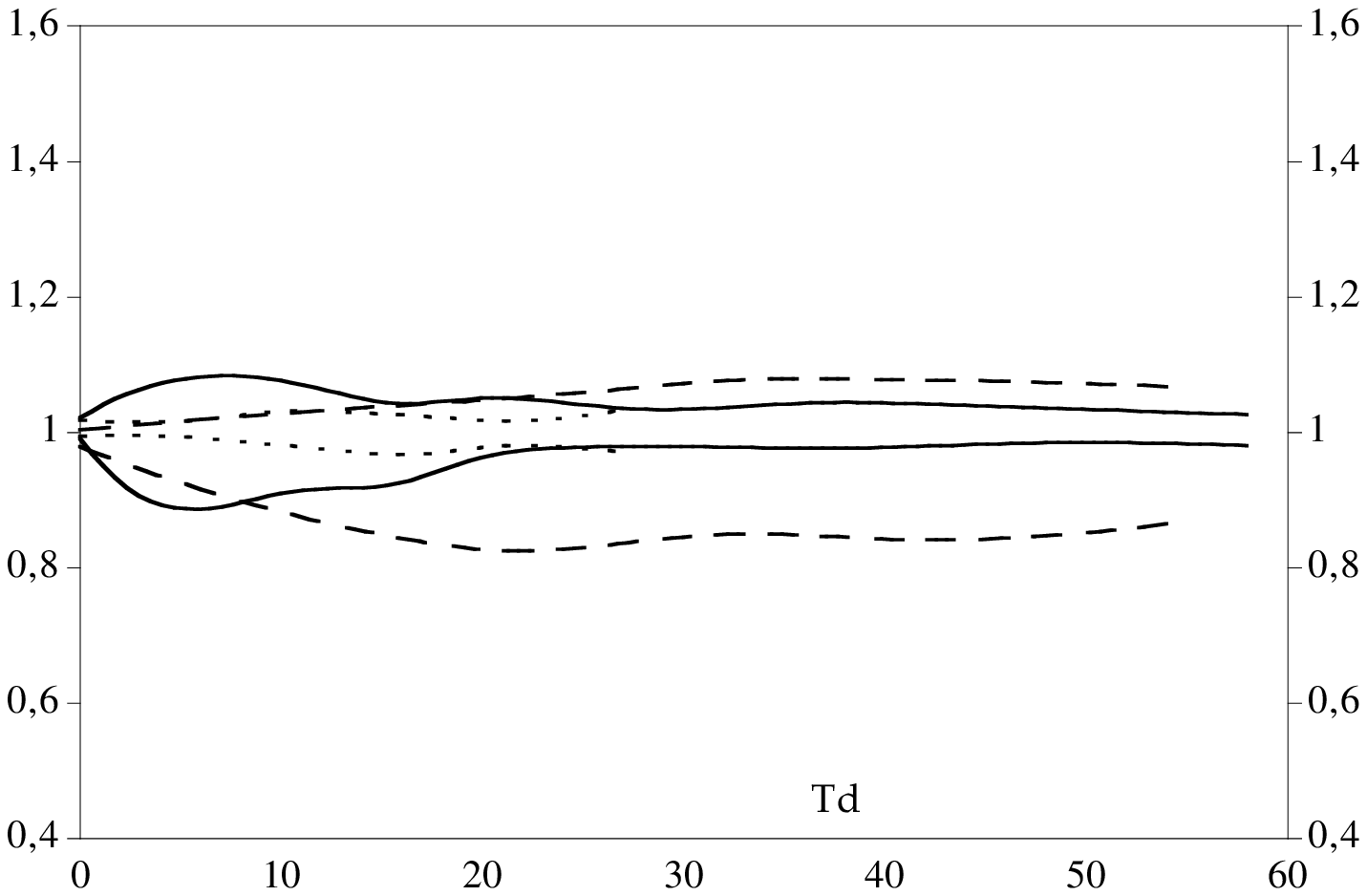,width=6cm}} \hfill\break \centerline{
\epsfig{file=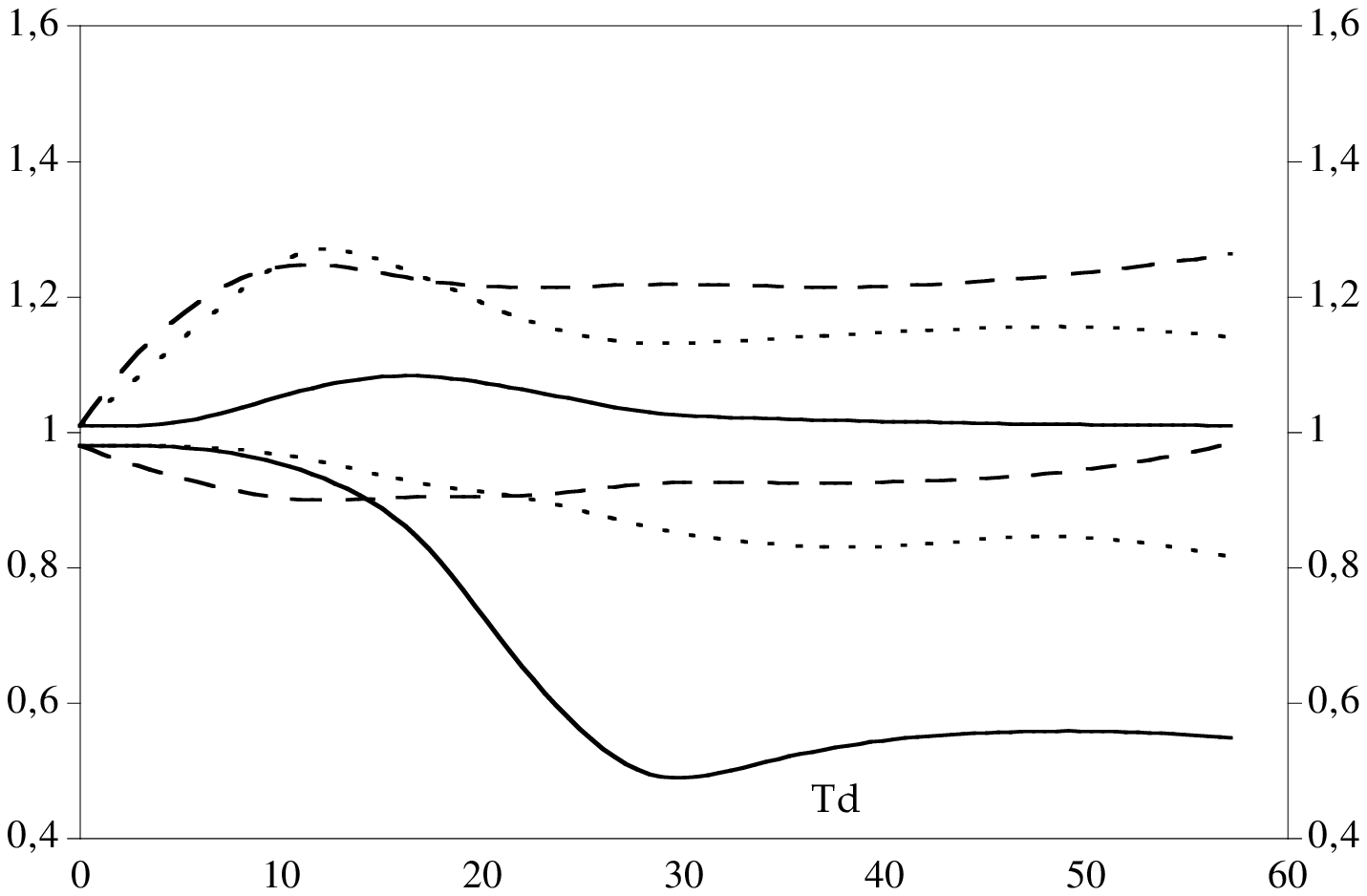,width=6cm}
\epsfig{file=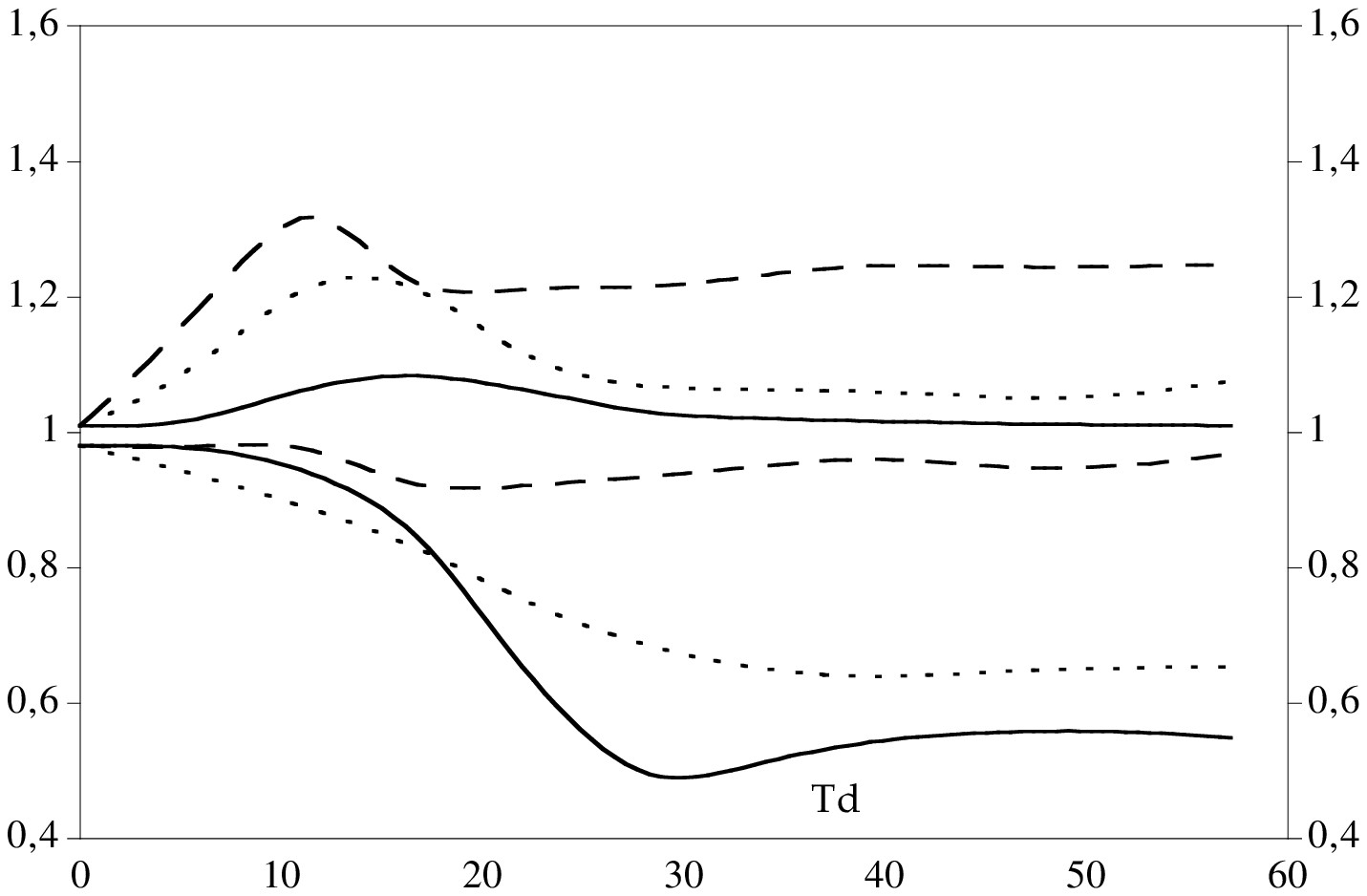,width=6cm}}
\caption{ The axial ratios vs.
 dynamical time for $n=4$; $r_a=100$-HR$_2$ (top-left panel), $r_a
 =100$-IR$_2$ (top-right panel), $r_a =1.5$-HR$_2$ (bottom-left
 panel) and $r_a= 1.5$-IR$_2$ (bottom-right panel). The
 amount of rotation is represented by using different kinds of lines:
 solid lines : $\tau=0$(HR) or $k=0$ (IR), dotted lines:
 $\tau=0.3$(HR) or $k=0.6$ (IR), and dashed lines: $\tau=0.8$(HR) or  
 $k=1.4$ (IR).
 \label{ratio2} } 
\end{figure*}

Let us  first consider rotating systems  with  high values  of $r_{\rm
  a}$.   We    recall that non-rotating    OM   systems with  the same
  anisotropy    parameter are stable   \cite{perez2}.   The  dynamical
  evolution obtained for such systems in  our present simulations (see
  Figure  \ref{ratio2}, top   panels)   allows us to   distinguish two
  classes of behavior.    When $r_{\rm a}$  is large  and the  rate of
  rotation stays modest (typically $\mu < 10\%$), we find that systems
  remain stable  and spherical ($a_{1}\simeq   a_{2}$, $f \simeq  1$).
  However, when $r_{\rm a}$ is large  and the rate of rotation becomes
  important (typically $\mu > 10\%$), we find that initially spherical
  systems develop a  very soft bar-like instability  ($a_{2}\simeq 1$,
  $a_{1}\simeq 0.85$, $f\simeq 1.3$).

Systems  with small  $r_{\rm   a}$ values  have  instead  a  different
behavior, which  depends   on whether the  rotational motion  has been
introduced by using  a homogeneous or  an inhomogeneous procedure.  In
the first case (HR$_{2}$  procedure), Figure \ref{ratio2} (bottom-left
panel) shows that    systems which are  radial  orbit-unstable without
rotation  (e.g., $a_{1}\simeq 0.55,  a_{2}\simeq 1$, $f \gg 1$) (solid
line), become quasi-spheroidal ($a_{1}\simeq 0.85$, $a_{2}\simeq 1.1$,
$f\simeq 1.3$) when they have a modest rotation  motion ($\mu < 10\%$)
(dotted line).  However, when  rotation is important (typically $\mu >
10\%$), such systems develop a disc-like instability ($a_{1}\simeq 1$,
$a_{2}\simeq 1.25$, $f\simeq 0$) (dashed  line).   In the second  case
(IR$_{2}$  procedure)(bottom-right panel), the  fact that rotation has
been introduced  only in a central region  of the  system prevents one
from     obtaining quasi-spherical   systems    and,  therefore,   the
radial-orbit  instability persists  ($a_{1}\simeq 0.65$,  $a_{2}\simeq
1.1$,  $f\simeq 3.5$) for systems  with  a  modest amount of  rotation
($n=4$;IR$_{2}$;$r_{\rm  a}=1.5$;$k=0.6$)  (dotted    line).  When the
amount of rotation is high enough ($\mu > 10\%$) a disc-like structure
appears ($a_{1}\simeq 1$, $a_{2}\simeq  1.25$, $f\simeq 0$), as in the
HR$_2$  procedure. The  evolution of the  axial  ratio for the  system
($n=4$; IR$_{2}$; $r_{a}=1.5$; $k=1.4$) is represented by dashed line.
These  results hold whatever the $n$  parameter value is.  In practice
we   have performed  numerical simulations  for   three values  of $n$
($n=3.5, 4, 4.5$).

\noindent

Are the stability  parameters  $P_{\epsilon}$ and $P_{\lambda}$  still
discriminating for such rotating systems?. We  can see in Table 2 that
non-rotating   stable systems ($r_{\rm  a}  =  100$)  are predicted to
remain   stable   when the      second rotation  method     is applied
($P_{\epsilon}$  stays smaller than 20  and $P_{\lambda}$ stays larger
than 2.5) whatever  the $\tau$ and  $k$ parameters values are.  On the
other  hand, non-rotating radial-orbit unstable  systems ($r_{\rm a} =
1.5$) are predicted to become stable when sufficient Method 2 rotation
is applied, from $\tau = 0.4$ and $k=0.8$  which correspond on figures
\ref{krothom}  and \ref{krotinh} to  typically  $\mu\simeq 5\%$. As  a
matter of fact, in this case, $P_{\epsilon}$ becomes  less than 20 and
$P_{\lambda}$ becomes greater than  2.5. Consequently, these stability
parameters which have been   constructed to predict the   stability of
non-rotating spherical   systems  are not  relevant   as soon   as the
quantity of kinetic rotation energy becomes  large, typically $\mu \ga
10 \%$.

\begin{table*}
\label{stabparam2}
\centerline{
\begin{tabular}{||c|cc|cc||c|cc|cc||} \hline
HR$_{2}$ & & & & & IR$_{2}$ & & & & \\
 $r_{\rm a}=100$ & & &$r_{\rm a}=1.5$ & & $r_{\rm a}=100$ &  \\ \hline
 $\tau $ &  $P_{\varepsilon}  $     & $P_{\lambda}$   &   $P_{\varepsilon}$  &
 $P_{\lambda}$   &  k  &   $P_{\varepsilon}    $  &  $P_{\lambda}$   &
 $P_{\varepsilon}$         &       $P_{\lambda}$    \\          \hline
 0.0&22.42&1.39&13.84&4.07&0.4&23.50&1.81&17.93&3.51\\
 0.2&21.31&1.94&13.43&4.08&0.6&21.68&2.17&13.34&3.57\\
 0.4&19.21&2.92&12.32&4.10&0.8&18.43&2.81&11.67&3.67\\
 0.6&16.96&3.55&10.46&4.17&1.0&16.56&3.17&10.55&3.75\\ 0.8&14.08&4.04&
 8.97&4.22&1.2&15.23&3.62&        9.39&3.80\\          1.0&11.24&4.58&
 7.40&4.46&1.4&13.85&3.97& 8.64&3.96\\ \hline
\end{tabular}
}
\caption{Evolution of the stability parameter for 
systems with $n=4$ and put in rotation by using the second method}
\end{table*}

\section{Physical Interpretation and Conclusions}

The  rotational  properties  of collapsed  systems  depend  to a large
extent on the amount of  angular moment before  the collapse. In order
to study in   a  realistic way the  importance  of  rotation   for the
dynamics    of self-gravitating systems,    it  is necessary either to
attempt  an  analytical approach, or to   perform a complete numerical
study modelling  the   collapse and  relaxation phases   prior to  the
two-body relaxation phase.  However, although the collapse of a system
can be studied  by using the  introduced amount  of rotational kinetic
energy as parameter, it is   difficult to extract general  conclusions
from this  kind of experiments.  As a matter  of fact  in this way the
post-collapse physical   features of the   object cannot be completely
controlled and hence it can   then be difficult   to study with  these
methods the influence of the rotation on post-collapse systems.

This justifies the method that  we have used in this
paper.  Starting from virialized systems  with exactly known dynamical
properties, we can study the  influence of rotation by controlling its
features.  If   the initial systems cover a   wide variety of physical
properties,  and if  the methods to  introduce   the rotation preserve
certain  fundamental features of    these systems (invariance of  mean
energy, conservation  or controlled  modification of the  distribution
function), the numerical study  will then be able to  be used as a model
to  extract  some general  conclusions.  As   a  matter  of fact,  our
simulations start  from a  wide  variety of  initial conditions  fully
controllable through the dependence    on the two parameters $n$   and
$r_a$. On the other hand, the techniques used to introduce rotation to
the  systems preserve (as explained in  Section 3) the properties that
ensure  that our   models  are spherical stationary   solutions of the
collisionless Boltzmann-Poisson system.

The main properties found in our study are the following :
\begin{itemize}
\item  There do  not   exist spherical self-gravitating systems   in
"fast" rotation. Our  simulations show in  fact that, when $\mu \ga 10
\%$, the systems do not remain spherical but become lengthened along
one  or   two  axes  depending  on  whether   they   are isotropic  or
anisotropic, respectively, when they do not have a rotational motion.
\item Rotation (in our case HR$_{2}$ and IR$_{2}$) can allow for a
reorganization of systems in  velocity space able to modify  their
dynamical behavior.  We have  in fact shown   that a moderate rotation
(typically  $0\le  \mu \la   10\%$)   can   stabilize and  confer    a
quasi-spherical structure to systems that, when they are not rotating,
suffer a radial-orbit instability. Therefore, there can exist rotating
spherical  self-gravitating systems.  This is  the case for our models
with $r_{\rm a} \ga 2$ and $\mu \la 10\%$.
\item  We have finally  found that the stability parameters introduced
  in \cite{perez2} remain valid as long as $\mu \la 10\%$.
\end{itemize}

\section{Appendix: Generation of initial conditions}

\subsection{The initial positions for the particles}
Let  us  consider a   density   $\rho_{iso}$ given by   the polytropic
relation $\rho_{iso}=c_{n}\psi_{iso}^{n}$, where
$$
c_n=\frac{(2\pi)^n
\Gamma(n-1/2)}{\Gamma(n+1)}\;\;\;\;,
\;\;\Gamma(n+1):=\int_{0}^{\infty}\!\!x^n
e^{-x}dx
$$
and $\psi_{iso}$ is the solution of the Lame-Emden equation
$$
\frac{1}{r^2}\frac{d}{dr}
\left(
r^2 \frac{d\psi}{dr}
\right)
+4\pi G c_n \psi^n =0
$$
This isotropic model is then deformed according to 
$$
\rho_{ani}(r)=(1+\frac{r^2}{r_{a}^2})\rho_{iso}
$$
where the anisotropic  radius $r_a$ is  a parameter which controls the
deformation. The polytropic index $n$ is chosen in the range $]0.5,5]$
in order to the system admits a finite mass $M(<r)=4\pi\int_{0}^{r}r^2
\rho_{ani}(r)dr$.  The total mass of the system is normalized to unity
and we then compute  for a large set   of particles ($1\leq i\leq  N$)
from the inverse function of $M(x)$ and $sin(x)$, the components
\begin{eqnarray}
r_{i} = r_{max} M^{-1}(x) \nonumber\\
\theta_{i}=2\arcsin(\sqrt{x}) \nonumber \\
\phi_{i}=2\pi.x \nonumber
\end{eqnarray}
where $x$ is  an uniform random variable on  $[0,1]$.  The size of the
system $r_{max}$ is chosen such that $\rho_{ani}(r>r_{max})<10^{-5}$.

\subsection{The  initial velocities for particles}
Let us consider the velocity components in spherical coordinates ($v_{r},
v_{\theta},            v_{\phi}$),              we             compute
$v_{t}=\sqrt{v_{\theta}^{2}+v_{\phi}^{2}}$       and           $\alpha
\;=\;\arctan(v_{\theta}/v_{\phi})$.  The probability  $p(\Gamma)$for
finding         a       particle          in            a       volume
$d\Gamma:=dr\;d\theta\;d\phi\;dv_{r}\;dv_{\theta} \; d\alpha$   of the
phase space is defined  from the DF of the system
\begin{eqnarray}
p(\Gamma)d\Gamma=\frac{1}{N}f(\Gamma)r^{2}dr\sin\theta d\theta   d\phi
v_{t}dv_{t} dv_{r}d\alpha \label{eqpro}
\end{eqnarray}
In the Ossipkov-Merritt model DF is  a function only on $Q$ variable 
\begin{eqnarray}
f=f(Q)\;\;\;\mbox{with}\;\;\;Q=\frac{1}{2}v_{r}^{2}+\frac{1}{2}\left(
1+\frac{r^{2}}{r_{a}^{2}}\right)v_{t}^{2}+\psi(r)
\end{eqnarray}
the equation (\ref{eqpro}) then reduces to  
\begin{eqnarray}
\int                   p(\Gamma)d\Gamma=\int            \frac{8\pi^{2}
r^{2}v_{t}}{N}f(r,v_{r},v_{t})dr\;dv_{t}\;dv_{r} \label{eqpro2}
\end{eqnarray}
$r$, $v_{r}$ and $v_{t}$ are dependant random variables.  In order
to continue the  integration  of equation (\ref{eqpro2}) we  introduce
the variables $R$ and $\beta$ defined as following
\begin{eqnarray}
v_{r}&=&R.\cos\beta               \nonumber                         \\
v_{t}\sqrt{1+\displaystyle{\frac{r^{2}}{r_{a}^{2}}}}&=&R.\sin\beta
\nonumber
\end{eqnarray}
We then get 
\begin{eqnarray}
f(r,v_{r},v_{t})   \frac{8\pi   r^{2}v_{t}}{N}= f\left(\frac{R^2}{2} +
\psi(r)\right)\frac{8\pi^{2}r^{2}R^{2}\sin\beta}{N\left
( 1+\displaystyle{\frac{r^{2}}{r_{a}^{2}}}\right)}drdRd\beta \nonumber
\end{eqnarray}
We see  from the previous  expression that the random variable $\beta$
is $r$ and $R$ independant and we have
\begin{eqnarray}
p(\beta)d\beta&=&\frac{\sin \beta}{2}d\beta \nonumber \\    p(r,R)drdR
&=&\frac{16
\pi^{2}r^{2}R^{2}}{1+\displaystyle{\frac{r^{2}}{r_{a}^{2}}}}f\left(
\frac{R^2}{2}+\psi(r)\right) drdR\nonumber
\end{eqnarray}
The conditional probablity  for finding a particle with a
velocity defined by $R$ at a given distance $r_0$ is then
\begin{eqnarray}
p(R\mid r_0)=\frac{p(r_0,R)}{p(r_0)}\;\;\;\;\;\mbox{with}\;\;\;\;\;
p(r_0)=\frac{4\pi r^{2}\rho(r_0)}{M(=1)}
\end{eqnarray}
and finally
\begin{eqnarray}
P(R\mid r=r_{o})&=&\frac{4\pi}
{\rho(r_{o})\left(1+\frac{r_{o}^{2}}{r_{a}^{2}}\right)}
\int_{0}^{R}R'^{2}f(\frac{R'^{2}}{2}+\psi(r_{o}))dR'\nonumber\\
&=&\frac{2\pi}
{\rho(r_{o})\left(1+\frac{r_{o}^{2}}{r_{a}^{2}}\right)}\int_{\psi(r_{o})}
^{\psi(r_{o})+\frac{R^{2}}{2}}\!\!\frac{f(Q)dQ}{\sqrt{2(Q-\psi(r_{o}))}}
\label{distriR} 
\end{eqnarray}
We   are now able to assign   a velocity for  each  particle which the
position have been  previously determined.
\begin{eqnarray}
R_{i} &=& P^{-1}(x|r=r_{o}) \nonumber \\
\beta_{i}&=&2\sin^{-1}(\sqrt{x})  \nonumber \\
\alpha_{i}&=&2\pi.x \nonumber
\end{eqnarray}
where $x$  is an uniform  random  variable on $[0,1]$,   and $P^{-1}$ is the
inverse    function   of    the  probability   defined   by   equation
(\ref{distriR}). Finally 
\begin{eqnarray}
v_{r\;i}&=&R_{i}.cos\beta_{i}                              \nonumber\\
v_{\theta\;i}&=&\sqrt{1+\frac{r_{o}^{2}}{r_{a}^{2}}}.R_{i}.\sin\beta_{i}\cos\alpha_{i}\nonumber\\
v_{\phi\;i}&=&\sqrt{1+\frac{r_{o}^{2}}{r_{a}^{2}}}.R_{i}.\sin\beta_{i}\sin\alpha_{i}\nonumber
\end{eqnarray}

\end{document}